\documentstyle[11pt]{article}

\renewcommand{\baselinestretch}{1}    
\begin{document}


\renewcommand{\baselinestretch}{1}
\thispagestyle{empty}	 

\title{
\bf Novel phases and finite-size scaling in\\
two-species asymmetric diffusive processes}
\author{\\ \\{
Kwan-tai Leung} \\ \\
{\small\it Institute of Physics, Academia Sinica},\\
{\small\it Taipei, Taiwan 11529, Republic of China} }
\vspace{1.2in}
\date{}        
\maketitle
\renewcommand{\baselinestretch}{1}

\centerline{ABSTRACT}
\medskip

We study a stochastic lattice gas of particles undergoing
asymmetric diffusion in two dimensions.
Transitions between a low-density 
uniform phase and high-density non-uniform phases characterized by
localized or extended structure are found.
We develop a mean-field theory which
relates coarse-grained parameters to microscopic ones. 
Detailed predictions for finite-size ($L$) scaling and density profiles
agree excellently with simulations.
Unusual large-$L$ behavior of the transition point 
parallel to that of self-organized sandpile models is found.

\medskip
\bigskip
\noindent{Pacs Numbers: 02.50.Ey, 02.60.Cb, 05.70.Fh, 64.60.Cn}

\newpage

There is much interest 
in the studies of lattice gas systems 
evolving under conservative stochastic 
dynamical rules\cite{spohnbook}.
When the rules do not satisfy detailed balance condition,
intriguing properties atypical
of thermal equilibrium arise\cite{SZ}.
Their mappings onto surface growth models
and directed polymer in a random medium\cite{mapping}
generates further interests and development. 
Very recently,  such connections have been exploited 
to shed light on the 
unbinding transitions of directed polymer\cite{krugtang}. 

One particularly simple class of lattice gas models 
consists of particles undergoing biased diffusion 
with no interaction except hard-core exclusion. 
Generally known as the asymmetric simple 
exclusion process (ASEP)\cite{spohnbook},
it has attracted the interests of both
physicists and mathematicians
for its novel phase transitions\cite{krugprl}
and shock structures\cite{shock} under various boundary conditions.
Generalizing the ASEP to a model 
having two species biased along 
{\em opposite\/} directions
in an $L_x \times L_y$ system,
Schmittmann {\em et al.\/}\cite{shz} observed
a transition between a uniform and a non-uniform phase 
as the number of particles is varied, with the latter phase characterized
by a compact strip of aggregated particles.
However, the fundamental question of the existence of the transition 
in the thermodynamic limit has not been answered.
This is not a simple issue to settle,
since the limiting process depends crucially on 
{\em anisotropic\/} finite-size effects
induced by making one direction special.
The precise form of anisotropies is unknown.
Nevertheless, 
a hint is provided by analogous continuous phase transitions,
which typically have ensemble averages depending on
$L_x^p/L_y$ with $p\ne 1$\cite{ktlprl}. 
Depending on how $L_x\to\infty$ and $L_y\to\infty$ are taken,
undesired singularities may arise from $L_x^p/L_y\to \infty$ or $0$.
If $p$ is not known,
it is not possible to proceed.
In another context,  
a cellular automaton model with {\em orthogonal\/} biases 
has been formulated 
to mimic cross traffic flow\cite{bihametal}.  
A similar jamming transition has also been found.
Whether the transition survives 
in the thermodynamic limit is likewise not investigated.

Motivated by these open questions, 
we study in this letter
a three-state model with two 
inter-penetrating ASEP orthogonal to each other.
By biasing the first (second) species along $+y$ ($+x$) direction,
{\em isotropic\/} finite-size effects are ensured
(i.e., effectively $p=1$ above)
so that no ambiguity is associated with
the thermodynamic limit,
whereas the essential simplicity of previous models is maintained.
Specifically,  we consider a model on a two-dimensional (2D)
square lattice of square geometry 
$L\times L$, with periodic boundary conditions. 
Occupation numbers $\{ n_{x,y} \}$ 
specify the configuration,
where $n_{x,y}\in \{0,1,2\}$ represents
a vacancy, a type-1 or a type-2 particle
at site $(x,y)$, respectively.  
Except hard-core exclusion,
the particles are non-interacting.
In simulations,
a site and one of its nearest neighbors are randomly picked
in turn.
If and only if one of them is vacant,
a jump occurs with rate:
$p$ for a type-1 particle hopping along $+y$, 
or a type-2 particle hopping along $+x$;  
$q$ for type-1 along $-y$, or type-2 along $-x$;  
and $r$ for type-1 along $\pm x$, or type-2 along $\pm y$.
$L^2$ such trials constitute one unit of time (one sweep).
Note that choosing $p>q\geq 0$ introduces the asymmetry, 
whereas $r>0$ provides transverse diffusions
necessary for ergodicity 
(contrary to the traffic flow automaton\cite{bihametal} which has $r=0$).
Using Metropolis rules and a parametrization by 
an ``electric field'' $\cal E$\cite{SZ}, the rates become
$p=1/4=r$, $q=e^{-\cal E}/4$. 

\underline{\sl Simulation results:}
Simulations are done on wide ranges of parameters:
$32\leq L\leq 256$, $0.2\leq {\cal E}\leq \infty$, with
equal density of particles, $\bar\rho_1=\bar\rho_2\equiv \bar\rho$.
While transitions between a uniform (U) phase and a strip (S)
phase are expected as $\bar\rho$ is varied,
a novel droplet phase (D) 
which is not present in previous models\cite{shz,bihametal}
apppears in between (see Fig.~1).
The three phases are characterized by different symmetries,
with the {\em localized\/} droplet breaking translational
invariance along both spatial directions and
drifting steadily forward along its symmetry axis.
That it is not 
an unstable or metastable state 
is revealed by the behavior of 
an order parameter, $\phi$
(essentially the Fourier amplitude 
of the density profile at a smallest wavevector\cite{ktlprl}),
which varies from one (compact strip) to
zero (perfectly uniform).
Fig.~1(a) depicts $\phi$ versus time ($t$) in the steady state
for $L=48$, ${\cal E}=\infty$,
showing frequent jumps among the three phases.
In fact, a histogram analysis shows that
with those parameters the system is at 
a three-phase coexistence,
i.e., at a {\em triple point\/}.
Deviating from this point, we find:
(a) 
At slightly different density, the U or S phase becomes dominant.
(b) 
For smaller bias $\cal E$ with fixed $L$,  
$\phi(t)$ plots and histograms show
that the triple point splits into two
transitions, e.g.,
S$\to$D at $\rho_{SD}$, then D$\to$U at $\rho_{DU}$,
as $\bar\rho$ decreases, 
with gradually widening 
gap ($\rho_{SD}-\rho_{DU}=0$---0.02 as $\cal E=\infty$---0.5, for $L=48$).
Diminishing first order characteristics --- 
hysterisis and finite jumps in $\phi$ and currents --- 
suggest that the transitions eventually turn continuous,
consistent with the observation that 
$\rho_{SD}$ grows with decreasing $\cal E$, 
making the droplet longer and fuzzier until no longer
distinquishable from a strip.
(c) 
For larger $L$ at fixed $\cal E$,
hysterisis becomes more pronounced while the transition points
shift systematically downward.

These results remind us of a pure substance
which has a triple point and 
a first order phase boundary ending at a critical point\cite{reichl}.
Another possibility is to have 
a first and a continuous order phase boundary joining at
a tri-critical point.
Despite the obvious interest, however, 
we have not been able to map out the phase diagram completely.
One difficulty is 
due to the above-mentioned decreasing transition densities
for increasing system sizes,
casting doubts on the survival
of the transitions as $L\to\infty$.
At the same time, growing hysterisis makes it harder
to locate the transitions precisely.
To understand these crucial finite-size effects,
we now focus our attention on a continuum theory.

\underline{\sl Mean-field theory:}
We begin by defining 
$P_{n n' n''\cdots}^y(x,y,t)$ as an equal-time, 
joint probability at time $t$ 
for the locate state at sites 
$(x,y)$, $(x,y+1)$, $(x,y+2),~\cdots$
to be $n$, $n'$, $n'', \cdots$ 
where $n=0,1$ or 2 as above 
($P_{n n' n''\cdots}^x$ likewise).
Since particles only hop to nearest neighbors,
it is easy to see, 
for $P_n(x,y,t+1)-P_n(x,y,t)\equiv \dot P_n$, that
\begin{eqnarray}
{1\over 2}\dot P_1 &=&
   p [ P_{10}^y(x,y-1)-P_{10}^y(x,y) ] 
  + q [ P_{01}^y(x,y)-P_{01}^y(x,y-1) ]  \nonumber  \\
 & & \mbox{}  + r [ P_{10}^x(x-1,y)-P_{10}^x(x,y) 
  + P_{01}^x(x,y)-P_{01}^x(x-1,y) ].
\label{P1dot}
\end{eqnarray}
The $1/2$ ensures the same time scale as in simulations,
and lattice constant$\equiv 1$.
$\dot P_2$ obeys a similar equation.
An attempt to close this set of equations 
generates a hierarchy in the usual way.
To proceed, we instead 
adopt a mean-field approximation in the following sense:
\begin{equation}
P_{nn'}^y(x,y) \approx P_n(x,y) P_{n'}(x,y+1)[ 1+\Delta_{nn'}^y(x,y)],
\label{MFT}
\end{equation}
with the asymmetric local correlations induced by biases accounted for
by $\Delta_{n n'}$'s (similarly for $P_{nn'}^x$).
Clearly $\Delta_{12}^y >0$, $\Delta_{21}^y <0$ etc.
Only those with $n\neq n'$ are kept,  as 
$\Delta_{nn} \ll 1$ is expected and 
well confirmed by simulation.

By taking the (naive) continuum limit 
of Eq.~(\ref{P1dot}) 
via Taylor expansions, 
assuming smooth functions,
we obtain from Eq.~(\ref{P1dot}) 
the continuity equation 
$\dot \rho_n = - \nabla \cdot {\bf J}_n$
for density $\rho_n$, with the current 
\begin{eqnarray}
{\bf J}_1 &=&
   \hat x E_{1x}\rho_1 \rho_2  
 + \hat y (E \rho_0 \rho_1 -  E_{1y}\rho_1 \rho_2 )  \nonumber \\
& & \mbox{} - \hat x D_\perp [(1-\rho_2) 
                  {\partial \rho_1\over \partial x}
                 + \rho_1 {\partial \rho_2\over \partial x} ]
 - \hat y D_\parallel [(1-\rho_2) 
                  {\partial \rho_1\over \partial y}
                 + \rho_1 {\partial \rho_2\over \partial y} ],
\label{rhodot}
\end{eqnarray}
where the parameters are completely specified
by the microscopics:
$E=2(p-q)$, $E_{1x}=2r(\Delta_{21}^x-\Delta_{12}^x)$,
$E_{1y}=2(p\Delta_{12}^y-q\Delta_{21}^y)$,
$D_\parallel=p+q$, and $D_\perp=2r$. 
${\bf J}_2$ can be obtained from 
${\bf J}_1$ by switching
the labels $1\leftrightarrow 2$, and
$x\leftrightarrow y$.
Before we continue, note that
the terms proportional to $E$'s and $D$'s represent physically drifts 
and diffusions respectively.
There is an induced drift normal to the bias 
in spite of symmetric transverse jump rates.
Similar but much weaker corrections 
in the diffusive pieces due to $\Delta$'s
are dropped in (\ref{rhodot}).
In the case of $p=q$, or $\rho_1=0$, or $\rho_2=0$, 
the stationary state is a uniform product measure 
and Eq.~(\ref{MFT}) becomes exact with $\Delta=0$.  Consequently,
it is necessary to have 
{\em two\/} species {\em driven\/} to 
produce all the nontrivial behavior.
Contrary to similar equations
derived from entropy consideration\cite{shz},
the microscopic origin of each term is made explicit.

Recall that there are three different steady states in simulations.
While the uniform solution is trivial to find,  
the droplet requires solving the full 2D problem.
Since the droplet phase occupies an extremely narrow range 
in $\bar\rho$ (i.e., $\rho_{SD}\approx \rho_{DU}$), 
for practical purposes we may focus on the uniform and the 
strip phase.
By setting ${\bf J}_n=$const, it is straightforward to
show that the strip profiles satisfy
\begin{eqnarray}
{1\over \varepsilon_0} \rho_1' &=&
\rho_1 (1-\rho_1+\rho_2) 
- {1\over \rho_0} ( \varepsilon_1 \rho_1\rho_2 + C )
\label{rho1prime}  \\
-{1\over \varepsilon_0} \rho_2' &=&
\rho_2 (1+\rho_1-\rho_2) 
- {1\over \rho_0} ( \varepsilon_1 \rho_1\rho_2 + C )
\label{rho2prime} 
\end{eqnarray}
for equal $\bar\rho_1=\bar\rho_2=\bar\rho$, 
where $\varepsilon_0=\sqrt{2}E/(D_\parallel+D_\perp)
=2\sqrt{2}(p-q)/(2r+p+q)$, 
$\varepsilon_1=(E_{1x}+E_{1y})/E=[p\Delta_{12}^y+r\Delta_{21}^x-
q\Delta_{21}^y-r\Delta_{12}^x]/(p-q)$, 
$\rho_n'\equiv d\rho_n / du$,  
with the $u$ axis perpendicular to the strip,
at a polar angle of $135^\circ$.
$C=\sqrt{2} {\cal J}/E$ is a reduced current along $+u$,
where ${\cal J}={\bf J}_1\cdot (\hat y-\hat x)/\sqrt{2}$.
$u$ and $\cal J$ have the same units as in simulations.

Using the data for $\Delta$'s, 
we find a remarkable result: 
$\varepsilon_1 \approx 1-2\bar\rho=\bar\rho_0$   
for {\em arbitrary\/} $\cal E$, $L$ and $\bar\rho$ in
the uniform phase (i.e., all data fall on one {\em universal line\/}).
While apparently 
$\Delta\to 0$ (so does $\varepsilon_1$)
as $\bar\rho_0\to 0$,
we have yet an explanation for the full, universal behavior.
In line with a mean-field approach,
we assume it remains to hold locally  
and replace $\varepsilon_1$ by $\rho_0(u)$ in 
Eqs.~(\ref{rho1prime}) and (\ref{rho2prime}).
Agreement between the predictions and simulations 
provides a {\em posteriori\/} justification\cite{footnotea}.
Then Eqs.~(\ref{rho1prime}) and (\ref{rho2prime}) contain
{\em no adjustable parameter\/} and 
imply the scaling form
\begin{equation}
\rho_n(u,{\cal J},L,{\cal E} )
=f_n(u\varepsilon_0, {\cal J}/E, L\varepsilon_0).
\label{profile}
\end{equation}
Unlikely to be obtained in closed forms,
the functions $f_n$'s  are determined by numerically solving 
the coupled Eqs.~(\ref{rho1prime}) and (\ref{rho2prime}).
One stringent test of the theory is to compare these functions 
with those in simulations.
Fig.~2 shows a typical test of not only 
$f_n$ but also the scaling in $L$.
With no adjustable parameter,  
the agreements are impressive verifications of our approach.
Deviation becomes noticeable only 
in the small $L\varepsilon_0$ regime.
Furthermore,  by (\ref{profile}), 
$(\sqrt{2}/L)\int du\, \rho_n(u)
=\bar\rho({\cal J}/E, L\varepsilon_0)$, 
or equivalently
\begin{equation}
C=\sqrt{2}{\cal J}/E =g(\bar\rho, L\varepsilon_0).
\label{jfss}
\end{equation}
This implies the entire current-density relation (along
with the transitions) is the same for different $L$
and microscopic $\cal E$, provided $L\varepsilon_0$ is fixed.
Fig.~3 confirms this prediction.
The $\bar\rho$ dependence of $g$ also compares 
favorably with simulations, except at small $L\varepsilon_0$.
This is consistent with 
the observed transitions gradually turning continuous 
at small biases,
as fluctuations ignored in the equations 
become significant.

Eq.~(\ref{jfss}) allows an investigation of the thermodynamic limit.
The transition point $\bar\rho_c(L,\cal E)$ 
is taken to be the minimum $\bar\rho$ allowed for the strip solution
of a given $L\varepsilon_0$ 
(i.e., at $dC/d{\bar\rho}=\infty$; cf. Fig.~3).
It is a function of $L\varepsilon_0$ alone.
$\lim_{L\varepsilon_0\to \infty}\bar\rho_c(L\varepsilon_0)$
then decides whether the transition survives or not.
Fig.~4 shows, for both simulations\cite{footnoteb} and theory,
that $\bar\rho_c \sim (L\varepsilon_0)^{-0.82(1)}$, 
and strongly suggests
$\bar\rho_c\to 0$ as $L\varepsilon_0\to \infty$. 

These results have far reaching implications.
If $\lim_{L\varepsilon_0\to \infty} \bar\rho_c=0$ indeed holds,
then {\em finite\/}-density transition 
will occur generically 
only for {\em infinitesimal bias\/} (i.e., ${\cal E} \propto  1/L$).  
It is hasty, however, to conclude that the
transition would actually be extinct for finite $\cal E$.
Seldom found in equilibrium,
this situation is by no means 
unfamiliar in non-equilibrium steady states.
A case of intense interest is the well studied 1D limited local 
sandpile models of self-organized
criticality\cite{kadanoffetal}.
Criticality of such 
models is controlled by the trough which has density
$\rho_t$ vanishing as $L^{-1/3}$\cite{CCGS},  
but the number of troughs $L \rho_t\to \infty$ 
as $L\to \infty$.
Analogously, 
our model has the strip width $\xi \sim L\bar\rho_c\to\infty$ 
but $\xi/L\to 0$ 
(hence $\xi$ is a {\em mesoscopic\/} length) 
as $L\to\infty$, 
so that, like the sandpiles, 
the transition remains well-defined for all $L$\cite{ktlSOCb}.   
We suspect 
other models have similar features,
but the corresponding analysis is hindered
by anisotropies\cite{shz} or by the lack of 
ergodicity\cite{bihametal}. 

To conclude,  
we have studied systems governed
by asymmetric diffusion and two-species exclusion 
as simple models exhibiting non-equilibrium phase transitions.
Our primary results are: 
(a) There exists multiple phases in spite of 
their apprarent simplicity;
(b) a quantitatively successful theory with known 
microscopic dependence has been developed, 
which could be improved to handle fluctuations 
by adding noise terms,  
and extended to more complicated models or related problems, such as 
with inter-particle interactions; and
(c) the survival of the associated 
transitions in the thermodynamic limit 
is addressed for the first time, and found
to be parallel to certain class of model sandpiles.
Finally, open questions such as 
the mechanism of stability for the droplet, 
details of the phase diagram and 
issues of critical exponents at small biases 
are of significant interests
to the general study of non-equilibrium phase transitions.

\underline{\sl Acknowledgments:}
The author thanks Prof. J.L. Lebowitz for an illuminating discussion,
and Dr. C.K. Chan for comments on the manuscript.
Support from the National Science Council of ROC under
grant number NSC83-0208-M-001-052 is gratefully acknowledged.

\bigskip
\underline{\sl Note added:}
At the completion of the manuscript,  we received a preprint
from Vilfan {\it et al.\/}\cite{vilfanetal},  
who obtained mean-field exact solutions 
for the strip phase in the oppositely
biased two-species model\cite{shz}.  
Their results agree qualitatively with ours, 
including a tentative tri-critical point,
but the large-$L$ limits have not been explored.


\newpage
\bigskip
\large{\bf Figure Captions}
\smallskip

\begin{enumerate}

\item 
(a) Order parameter at three-phase coexistence
for $L=48$, ${\cal E}=\infty$, $\bar\rho=0.1623$:
large, intermediate and small $\phi$ value corresponds to 
strip, droplet and uniform phase, respectively;
(b) a typical strip for $L=128$, $\bar\rho=0.076$; and
(c) a typical droplet for $L=128$, $\bar\rho=0.074$.
Light (dark) particles drift to the right (top).

\item 
Finite-size scaling of density profiles at
fixed $L\varepsilon_0=23.65$ (or fixed $C$) and $\bar\rho=0.36$:  
$L=32(\Box)$, 48$(\bigtriangleup)$, 64$(\circ)$, 
128$(\times)$, 256(dashed line), and 
theory (solid line) 
[cf. Eq.~(\ref{profile})] with no
free parameter.

\item 
Scaling of current
[Eq.~(\ref{jfss})].
Thin line is $\bar\rho(1-\bar\rho)(1-2\bar\rho)$ 
for uniform phase [from (\ref{rho1prime})].
Branches are for strip phase at various $L\varepsilon_0$:
(a) 6.445, $L=48(\circ)$; 
(b) 14.81, $L=32(\Box)$, 48$(\bigtriangleup)$, 64$(\bigtriangledown)$;  
(c) 23.65, 
$L=48(\bullet)$, 64$(\bigtriangleup)$, 128$(\Box)$, 256$(\times)$;
(d) 33.84, $L=48(\bigtriangledown)$; 
(e) 45.26, $L=48(\Box)$;
(f) 75.43, $L=80(\bigtriangledown)$.
Note confirmed scaling in (b) and (c).
Heavy lines are theoretical predictions for (b) to (f).

\item 
Large-$L$ behavior of 
transition density:
theory[cf. Fig.~3]($\bigtriangleup$); 
data($\bullet$ with error bars);
linear fit(dotted line).

\end{enumerate}


\begin{thebibliography}{99}
\bibitem{footnoteb}
For $L\varepsilon_0 > 45$,  $\rho_{SD}$, $\rho_{DU}$ and their
counterparts for reverse transitions begin to deviate.  Since our
predictions come from the strip solution,  
$\rho_{SD}(\approx\rho_{DU}$ for all $L\varepsilon_0)$
is used to compare with theory.
\end{thebibliography}
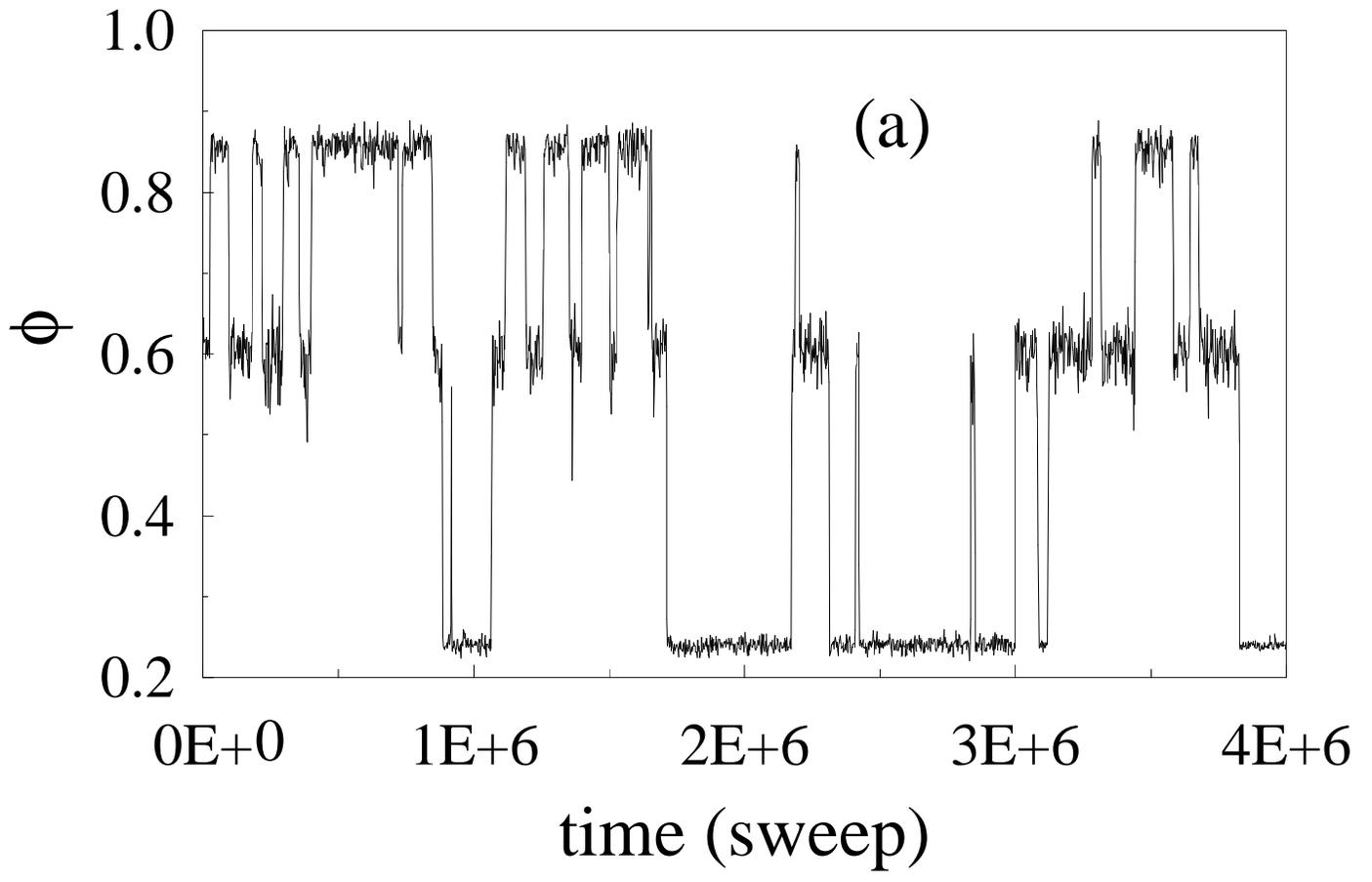
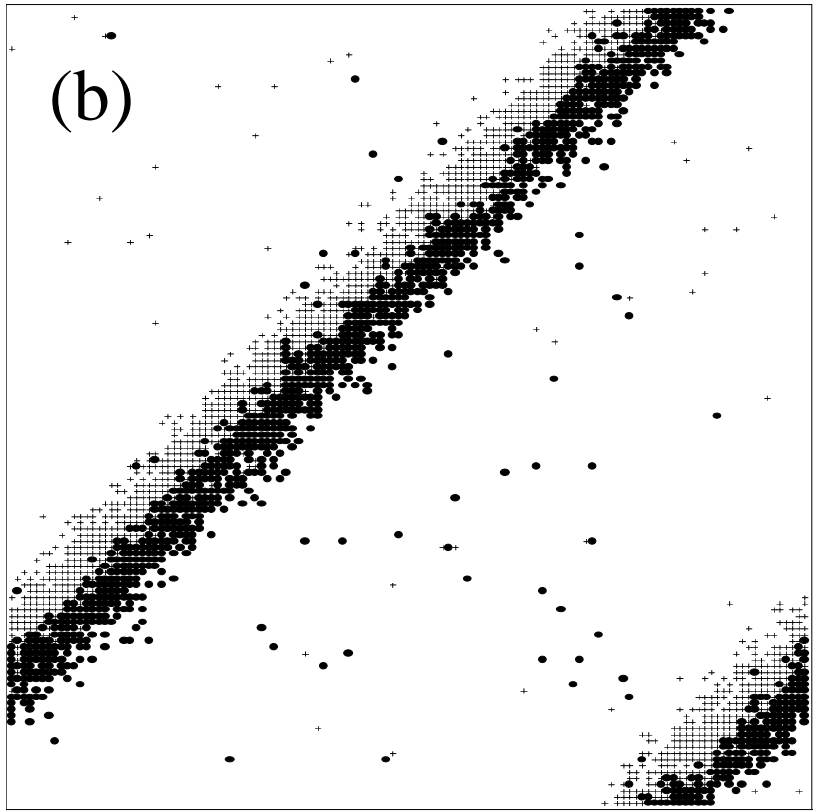
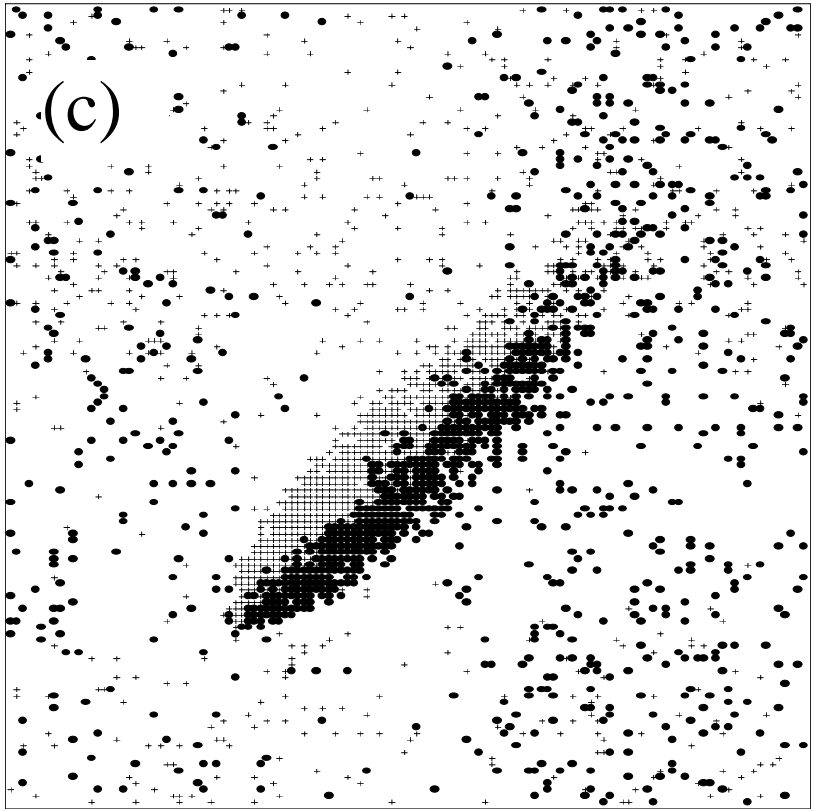
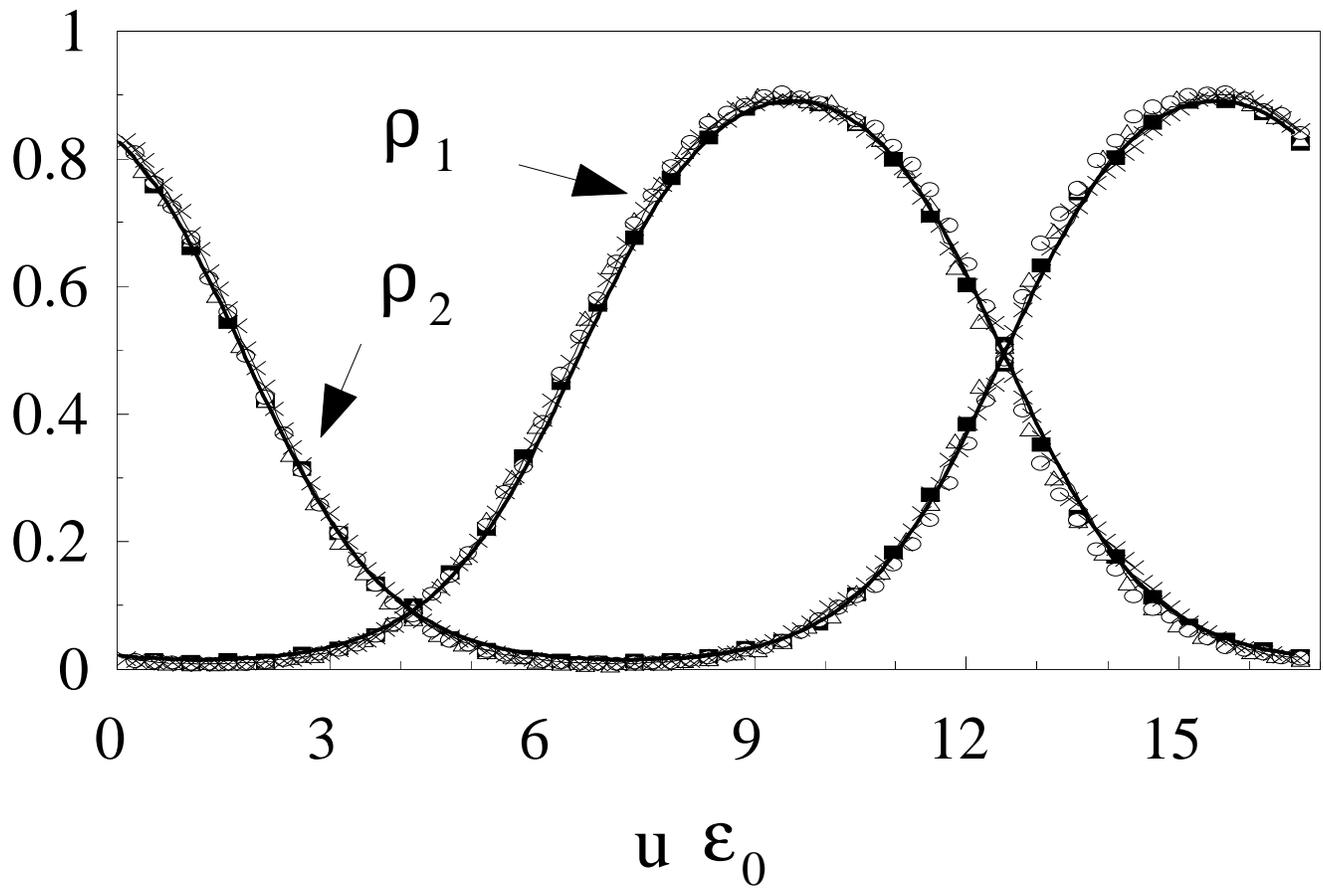
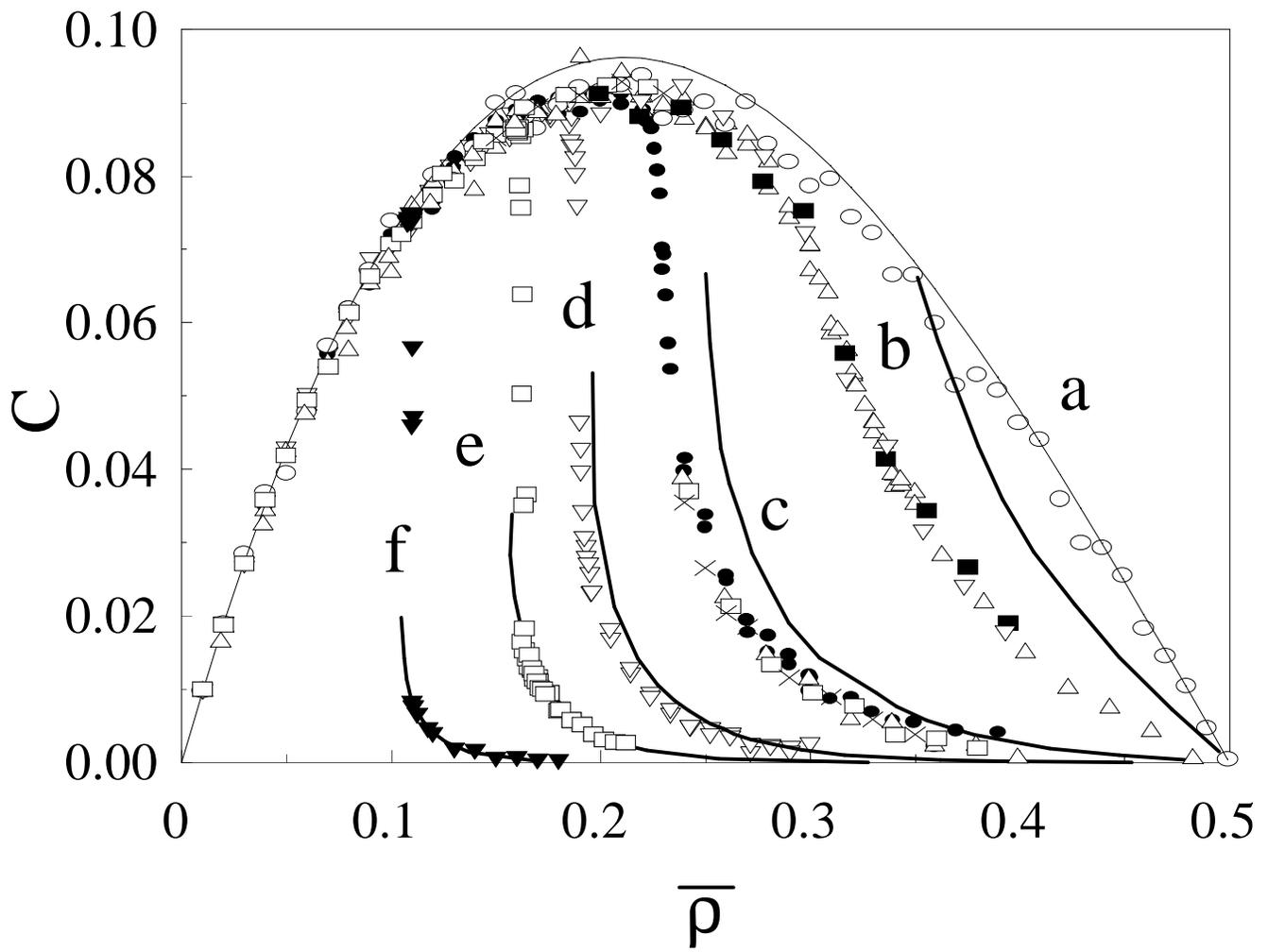
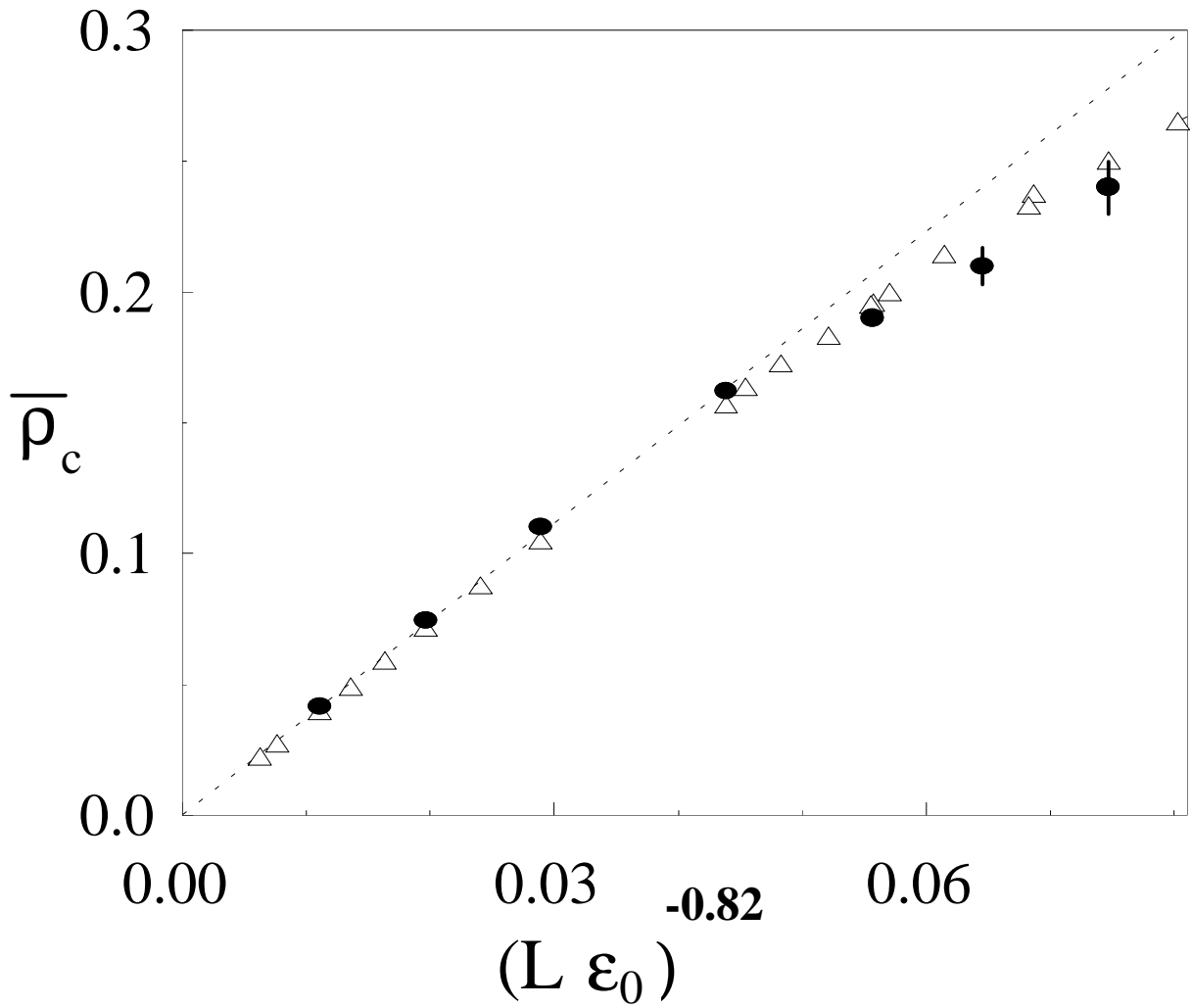
\end{document}